\documentclass[fleqn,twoside]{article}
\usepackage{espcrc2}
\usepackage{epsfig}

\title{Precision measurement of the hadronic cross-section 
through the radiative return method}

\author{Germ\'an Rodrigo~\address{Theory Division, CERN, 
        CH-1211 Geneva 23, Switzerland.}%
        \thanks{Supported in part by E.U. TMR grant HPMF-CT-2000-00989; 
        e-mail: {\tt german.rodrigo@cern.ch}},
        Henryk Czy\.z~\address{Institute of Physics, University of Silesia,
        PL-40007 Katowice, Poland.}%
	\thanks{Supported in part by EC 5-th Framework, 
        contract HPRN-CT-2000-00149; e-mail: {\tt czyz@us.edu.pl}}
        and
        Johann H. K\"uhn~\address{Institut f\"ur Theoretische Teilchenphysik,
        Universit\"at Karlsruhe, D-76128 Karlsruhe, Germany.}%
	\thanks{e-mail: {\tt jk@particle.uni-karlsruhe.de}}}
       
\begin{document}

\begin{abstract}
Electron--positron annihilation into hadrons plus an
energetic photon from initial-state radiation allows the hadronic
cross-section to be measured over a wide range of energies
at high luminosity meson factories. Weighted integrals over this
cross-section are a decisive input for electroweak precision tests.
A Monte Carlo event generator called PHOKHARA has been developed, which 
simulates $e^+ e^- \to$ hadrons + photon(s) at the NLO accuracy.
The latest tests and upgrades are presented in this paper. 
\end{abstract}

\maketitle

\section{INTRODUCTION}

The total cross-section for electron--positron annihilation into
hadrons is one of the fundamental observables in particle physics. Its
high energy behaviour provides one of the first and still most
convincing arguments for the point-like nature of quarks. Its
normalization was evidence for the existence of quarks of three
different colours, and the recent precise measurements even allow
for an excellent determination of the strong coupling at very high 
\cite{:2001xv} and intermediate energies 
(e.g. \cite{Kuhn:2001dm} and refs. therein) 
through the influence of QCD corrections.

Weighted integrals over this cross-section with properly chosen kernels
are, furthermore, a decisive input for electroweak precision tests.
This applies, for example, to the electromagnetic coupling at higher
energies or to the anomalous magnetic moment of the muon.

Of remarkable importance for these two applications is the low energy region,
say from threshold up to centre-of-mass system (cms) energies of approximately 
3~GeV and 10~GeV, respectively. Recent measurements based on energy scans 
between 2 and 5~GeV have improved the accuracy in part of this
range. However, similar, or even further improvements below 2~GeV
would be highly welcome. The region between 1.4~GeV and 2~GeV, in
particular, is poorly studied and no collider will cover this region 
in the near future. Improvements or even an independent cross-check 
of the precise measurements of the pion form factor in the low 
energy region by the CMD2 and DM2 collaborations 
would be extremely useful, in particular in view of the 
disagreement found with the analysis based on isospin-breaking-corrected 
$\tau$ decays~\cite{Davier:2002dy}. Since this dominates in
the analysis of the muon anomalous magnetic moment, this kind of studies
will help to clarify the situation with respect to the experimental 
measurement~\cite{Bennett:2002jb} of this quantity.

\section{RADIATIVE RETURN AT MESON FACTORIES}

Experiments at present electron--positron colliders operate mostly at 
fixed energies, albeit with enormous luminosity, with BABAR
and BELLE at 10.6~GeV, CLEO-C in the region between 3 and 5~GeV,
and KLOE at 1.02 GeV as most prominent examples. 

This peculiar feature allows the use of the radiative return, i.e. the reaction
\begin{equation}
e^+(p_1)+ e^- (p_2) \to \gamma(k_1) + \gamma^*(Q) (\to \mathrm{hadrons})~,
\label{eq:reaction}
\end{equation}
to explore a wide range of $Q^2$ in a single 
experiment~\cite{Binner:1999bt,Melnikov:2000gs,Czyz:2000wh,Spagnolo:1999mt,Khoze:2001fs,Hoefer:2001mx}. 

Nominally an invariant mass of the hadronic system between 
$2 m_\pi$ and the cms energy of the experiment is accessible. 
In practice, to clearly identify 
the reaction, it is useful to consider only events with a hard photon 
--- tagged or untagged --- which lowers the energy significantly.

The study  of events with photons emitted under both large and small 
angles, and thus at a significantly enhanced rate, is particularly 
attractive for the $\pi^+\pi^-$ final state with its clear signature, 
an investigation performed at present at 
DA$\Phi$NE~\cite{lastkloe,Aloisio:2001xq,Denig:2001ra,Adinolfi:2000fv}.
Events with a tagged photon, emitted under a large angle with
respect to the beam, have a clear signature and are thus particularly
suited to the analysis of hadronic final states of higher
multiplicity~\cite{babar}.

\section{MONTE CARLO SIMULATION}

To arrive at reliable predictions including kinematical cuts as 
employed by realistic experiments, a Monte Carlo generator 
is indispensable. The inclusion of radiative corrections 
in the generator and the analysis
is essential for the precise extraction of the cross-section. 
For hadronic states with invariant masses below 2 or even 3~GeV, 
it is desirable to simulate the individual exclusive channels with 
two, three, and up to six mesons, i.e. pions, kaons, etas, etc.,
which requires a fairly detailed parametrization of the various form 
factors. 

A first program, called EVA, was constructed
some time ago~\cite{Binner:1999bt} to simulate the production of a 
pair of pions together with a hard photon. 
It includes initial-state radiation (ISR), final-state
radiation (FSR), their interference, and the dominant radiative
corrections from additional collinear radiation 
through structure function (SF) techniques~\cite{Caffo:1997yy}.
This project was continued with the construction of a 
generator for the radiative production of four pions~\cite{Czyz:2000wh}.  

As a further development of this project a new Monte Carlo generator 
called PHOKHARA~\cite{Rodrigo:2001kf} has been constructed, which 
includes, in contrast to the former generators, the complete 
next-to-leading order (NLO) radiative corrections.
The first version of PHOKHARA incorporates ISR only and is limited to 
$\pi^+ \pi^- \gamma (\gamma)$ and $\mu^+ \mu^- \gamma (\gamma)$
as final states. PHOKHARA exhibits a modular structure that 
simplifies the implementation of additional hadronic modes
or the replacement of the currents(s) of the existing modes. 

In this paper we show how PHOKHARA is constructed and present some results 
obtained with its current version, as well as a comparison with 
the other aforementioned generators. Finally, 
the new features that will be part of the next version
of PHOKHARA are outlined and some preliminary results are presented. 
Further details will be given, however, in a forthcoming
publication~\cite{inpreparation}. 
These programs and the future versions of PHOKHARA can be 
downloaded from {\tt http://cern.ch/german.rodrigo/phokhara/}.

\section{NLO CORRECTIONS TO ISR}

\begin{figure}
\begin{center}
\epsfig{file=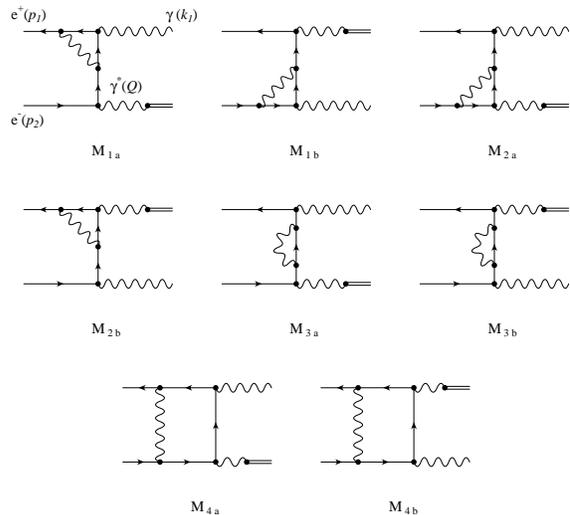,width=7.5cm} \vspace{-1.5cm}
\end{center}
\caption{One-loop corrections to initial-state radiation in 
$e^+ e^- \rightarrow \gamma +$ hadrons.}
\label{fig:nlo}
\end{figure}

\begin{figure}
\begin{center}
\epsfig{file=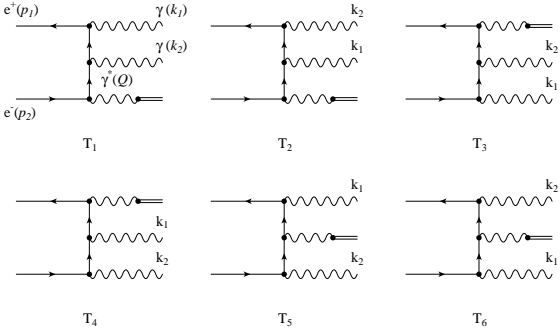,width=7.5cm} \vspace{-1.5cm}
\caption{Emission of two real photons from the initial state 
in $e^+e^-$ annihilation into hadrons.}
\label{fig:real}
\end{center}
\end{figure}

At NLO, the $e^+ e^-$ annihilation process~(\ref{eq:reaction}),
where the virtual photon converts into a hadronic final state,
$\gamma^*(Q) \rightarrow$ hadrons, and the real one is emitted
from the initial state, receives contributions from one-loop 
corrections (see Fig.~\ref{fig:nlo}) and from the 
emission of a second real photon (see Fig.~\ref{fig:real}). 

After renormalization the one-loop matrix elements still contain  
infrared divergences. These are cancelled by adding the contribution
where a second photon has been emitted from the initial state.
This rate is integrated analytically in phase space 
up to an energy cutoff $E_{\gamma}<w\sqrt{s}$ far below $\sqrt{s}$. 
The sum is finite; however, it depends 
on this soft photon cutoff. The contribution from the emission of 
a second photon with energy $E_{\gamma}>w\sqrt{s}$ completes the 
calculation and cancels this dependence. 

In order to facilitate the extension of the Monte Carlo simulation 
to different hadronic exclusive channels, the differential rate for the 
virtual and soft QED corrections is cast into the product of a leptonic 
and a hadronic tensor and the corresponding factorized phase space:
\begin{eqnarray}
d\sigma &=& \frac{1}{2s} \ L_{\mu \nu} \ 
d \Phi_2(p_1,p_2;Q,k_1) \nonumber \\ 
&\times& H^{\mu \nu} \ d \Phi_n(Q;q_1,\cdot,q_n) \
\frac{dQ^2}{2\pi}~,
\end{eqnarray}
where $d \Phi_n(Q;q_1,\cdot,q_n)$ denotes the hadronic 
$n$-body phase space including all statistical factors 
and $Q^2$ is the invariant mass of the hadronic system.

The physics of the hadronic system, whose description is 
model-dependent, enters only through the hadronic tensor 
\begin{equation}
H^{\mu \nu} = J^{\mu} J^{\nu +}~,
\end{equation}
where the hadronic current has to be parametrized through form 
factors~\cite{Czyz:2000wh,Kuhn:1990ad,Decker:1994af,Ecker:2002cw}.

The leptonic tensor, which describes the NLO 
virtual and soft QED corrections to initial-state radiation in 
$e^+ e^-$ annihilation, has the following general form: 
\begin{eqnarray}
L^{\mu \nu} &=&
\frac{(4 \pi \alpha)^2}{Q^4} \;
\bigg[ a_{00} \; g^{\mu \nu} + a_{11} \; \frac{p_1^{\mu} p_1^{\nu}}{s} 
\nonumber \\ &+& a_{22} \; \frac{p_2^{\mu} p_2^{\nu}}{s} 
 +  a_{12} \; \frac{p_1^{\mu} p_2^{\nu} + p_2^{\mu} p_1^{\nu}}{s} 
\nonumber \\ &+& i \pi \; a_{-1} \; 
\frac{p_1^{\mu} p_2^{\nu} - p_2^{\mu} p_1^{\nu}}{s} \bigg]~.
\label{generaltensor}
\end{eqnarray}
Terms proportional to $Q^{\mu}$ are absent as a consequence of 
current conservation. 
The scalar coefficients $a_{ij}$ and $a_{-1}$ allow the following expansion 
\begin{equation}
a_{ij} = a_{ij}^{(0)} + \frac{\alpha}{\pi} \; a_{ij}^{(1)}~, \qquad
a_{-1} = \frac{\alpha}{\pi} \; a_{-1}^{(1)}~,
\label{eq:expand}
\end{equation}
where $a_{ij}^{(0)}$ provide the leading order (LO) contribution 
and the imaginary antisymmetric piece proportional to $a_{-1}$ appears 
for the first time at NLO. 

As an alternative one can replace the Cartesian basis (\ref{generaltensor})
by a basis derived from the three circular polarization vectors
of the virtual photon $\varepsilon_L$ and $\varepsilon_\pm$:
\begin{equation}
L^{\mu \nu} =  \frac{(4 \pi \alpha)^2}{Q^4} \;
\sum a_{ij} \; \varepsilon_i^{*\mu} \varepsilon_j^{\nu}~, \quad
i,j=L,\pm~,
\end{equation}
where only four of the scalar coefficients are independent 
\begin{eqnarray}
 a_{L -} = a_{L +}~, \qquad 
& & a_{- L} = a_{+ L} = a_{L +}^*~, \nonumber \\
 a_{--} = a_{+ +}~, \qquad 
& & a_{- +} = a_{+ -}~. \nonumber 
\end{eqnarray}
An expansion similar to (\ref{eq:expand}) holds for these scalar coefficients.
The relationship between the components in both bases as well as expressions 
for these coefficients can be found in~\cite{Rodrigo:2001jr,Kuhn:2002xg}. 
The trace of the leptonic tensor,
which is related to the cross-section after angular averaging of the 
hadronic tensor, is particularly simple in the second case:
\begin{equation}
L^{\mu \nu} (Q_{\mu} Q_{\nu}-g_{\mu \nu} Q^2) = 
\frac{(4 \pi \alpha)^2}{Q^2} ( a_{LL} + 2a_{+ +} )~.
\end{equation}

Note that the imaginary part of $L_{\mu \nu}$, which is present
in the coefficients $a_{L+}$ or $a_{-1}$ only, is of interest for 
those cases where the hadronic current receives contributions from 
different amplitudes with non-trivial relative phases. This is 
possible, e.g. for final states with three or more mesons or for 
$p\bar{p}$ production.

The matrix elements for the emission from the initial state
of two real hard photons, i.e. $E_{\gamma}>w\sqrt{s}$, 
\begin{equation}
e^+(p_1) + e^-(p_2) \rightarrow  \gamma^*(Q) + \gamma(k_1) + \gamma(k_2)~,
\end{equation}
are calculated numerically following 
the helicity-amplitude method with the conventions 
introduced in~\cite{Jegerlehner:2000wu,Kolodziej:1991pk}.
As a test, the square of this matrix element averaged over initial 
particle polarization has also been calculated using the standard 
trace technique and tested numerically against the helicity method 
result. 

The virtual plus soft contribution and the hard one depend 
separately on the soft photon cutoff $w$ used to regulate the 
infrared divergences of the virtual diagrams.
The former shows a logarithmic $w$ dependence. 
The second, after numerical integration in phase space, exhibits 
the same behaviour, whereas their sum must be independent of $w$. 
To explicitly demonstrate this $w$-independence is therefore a basic
test of the program. Then the value of $w$ that optimizes the event 
generation, avoiding at the same time the appearance of unphysical 
negative weights, is determined.

Table~\ref{tab:epstest} presents the total cross-section 
for radiative production of a pair of pions calculated 
for several values of the soft photon cutoff at three different cms 
energies for the kinematical cuts from Table~\ref{tab:cuts}. 
The excellent agreement, within the error of the numerical integration, 
confirms the $w$-independence of the result. 
A value around $w=10^{-4}$ seems to be the best choice~\cite{Rodrigo:2001kf}.
 
\begin{table*}
\caption{Total cross-section (nb) for the process 
$e^+ e^- \rightarrow \pi^+ \pi^- \gamma$ at NLO for different values 
of the soft photon cutoff. Only initial-state radiation.
Cuts from Table~\ref{tab:cuts}.}
\label{tab:epstest}
\begin{center}
\begin{tabular}{cccc}
$w$ & $\sqrt{s}=$1.02~GeV & 4~GeV & 10.6~GeV \\ \hline 
$10^{-3}$ & 2.0324 (4) & 0.12524 (5) & 0.010564 (4)\\
$10^{-4}$ & 2.0332 (5) & 0.12526 (5) & 0.010565 (4)\\
$10^{-5}$ & 2.0333 (5) & 0.12527 (5) & 0.010565 (5)\\ \hline
\end{tabular}
\end{center}
\end{table*}

\begin{table*}
\caption{Kinematical cuts applied at different cms energies:
minimal energy of the tagged photon ($E_{\gamma}^\mathrm{min}$), angular 
cuts on the tagged photon ($\theta_{\gamma}$) and the pions ($\theta_{\pi}$),
and minimal invariant mass of the hadrons plus the tagged photon
($M^2_{\pi^+\pi^-\gamma}$)}
\label{tab:cuts}
\begin{center}
\begin{tabular}{cccc}
                   & $\sqrt{s}=$1.02~GeV & 4~GeV & 10.6~GeV \\ \hline  
$E_{\gamma}^\mathrm{min}$ (GeV) & $0.01$ & $0.1$ & $1$  \\
$\theta_{\gamma}$ (degrees)  & $[5,21]$   & $[10,170]$ & $[25,155]$ \\
$\theta_{\pi}$ (degrees)     & $[55,125]$ & $[20,160]$ & $[30,150]$ \\
$M^2_{\pi^+\pi^-\gamma}$ (GeV$^2$) & $0.9$ & $12$ & $90$ \\ 
\hline
\end{tabular}
\end{center}
\end{table*}

\begin{table*}
\caption{Total cross-section (nb) for the process 
$e^+ e^- \rightarrow \pi^+ \pi^- \gamma$ at LO, NLO and in the 
collinear approximation via structure functions (SF), with the cuts from 
Table~\ref{tab:cuts}. Only initial-state radiation. NLO(2) gives 
the NLO result with the same cuts as NLO(1) but for the hadron--photon 
invariant mass unbounded.}
\label{tab:xsec}
\begin{center}
\begin{tabular}{lccc}
  & $\sqrt{s}=$1.02~GeV & 4~GeV & 10.6~GeV \\ \hline  
Born     & 2.1361 (4) & 0.12979 (3) & 0.011350 (3) \\
SF       & 2.0192 (4) & 0.12439 (5) & 0.010526 (3) \\
NLO (1)  & 2.0332 (5) & 0.12526 (5) & 0.010565 (4) \\ 
NLO (2)  & 2.4126 (7) & 0.14891 (9) & 0.012158 (9) \\ 
\hline
\end{tabular}
\end{center}
\end{table*}

\begin{table*}
\caption{Total cross-section (nb) for initial-state radiation in 
the process $e^+ e^- \rightarrow \mu^+ \mu^- \gamma$ at LO,
NLO (1) and NLO (2) with the cuts from Table~\ref{tab:cuts}, 
the pions being replaced by muons.}
\label{tab:mxsec}
\begin{center}
\begin{tabular}{lccc}
  & $\sqrt{s}=$1.02~GeV & 4~GeV & 10.6~GeV \\ \hline  
Born     &  0.8243(5) &  0.4690(6) &  0.003088(6) \\
NLO (1)  &  0.7587(5) &  0.4449(6) &  0.002865(6) \\ 
NLO (2)  &  0.8338(7) &  0.4874(14) &  0.00321(6) \\ 
\hline
\end{tabular}
\end{center}
\end{table*}

\begin{table}
\caption{Total cross-section (nb) for the process 
$e^+ e^- \rightarrow \pi^+ \pi^- \gamma$ at $\sqrt{s}=$1.02~GeV in NLO 
and in the collinear approximation (SF) as a function of the cut on the 
invariant mass of the hadron + tagged photon $M^2_{\pi^+ \pi^- \gamma}$.
Only initial-state radiation. Minimal energy of the tagged photon and 
angular cuts from Table~\ref{tab:cuts}.}
\label{tab:invariantcut}
\begin{center}
\begin{tabular}{ccc}
$M^2_{\pi^+ \pi^- \gamma}$ (GeV$^2$) & SF & NLO \\ \hline  
0.1  & 2.4127(18) & 2.4132(8)\\
0.2  & 2.4126(18) & 2.4131(8)\\
0.3  & 2.4124(18) & 2.4127(8)\\
0.4  & 2.4098(18) & 2.4096(8)\\
0.5  & 2.3949(18) & 2.3953(8)\\
0.6  & 2.3425(16) & 2.3455(8)\\
0.7  & 2.2449(11) & 2.2543(8)\\
0.8  & 2.1387(9) & 2.1533(8)\\
0.9  & 2.0198(8) & 2.0334(8)\\
0.95 & 1.9437(8) & 1.9522(8)\\
0.99 & 1.8573(8) & 1.8559(8)\\
\hline
\end{tabular}
\end{center}
\end{table}

\section{LL VERSUS NLO}

\begin{figure}
\begin{center} 
\vspace{-.5cm}
\epsfig{file=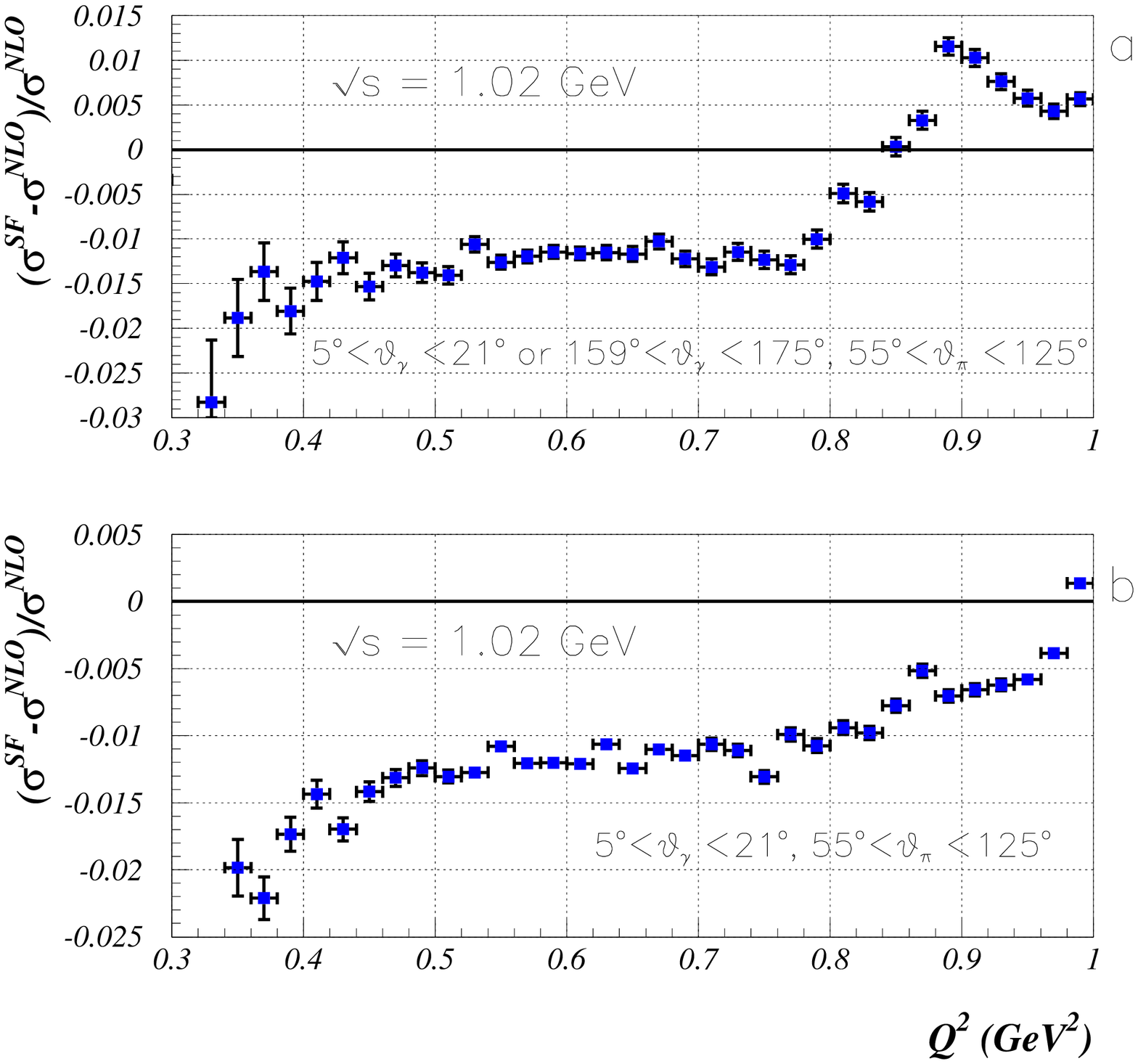,width=7.5cm} 

\epsfig{file=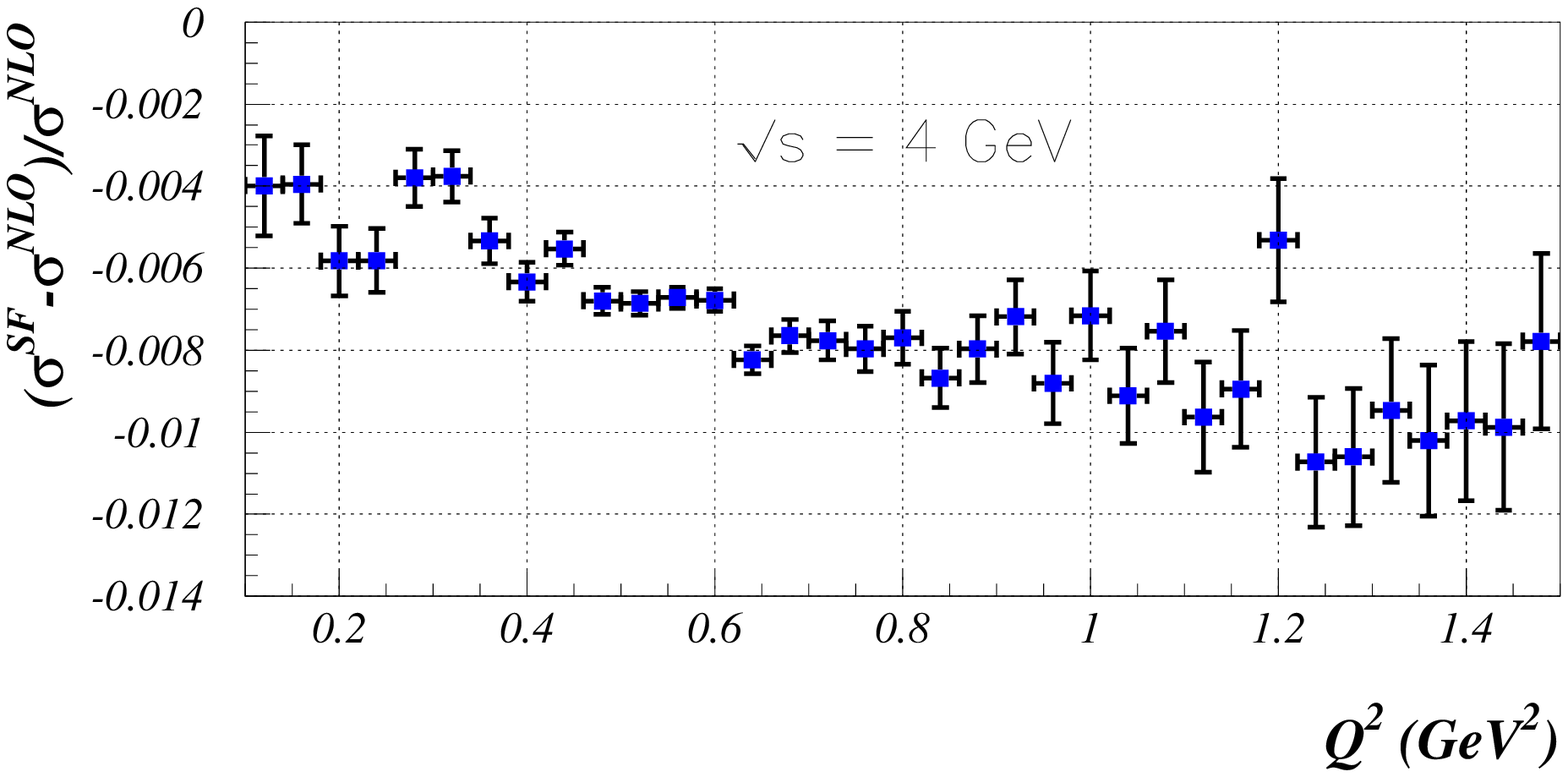,width=7.5cm} 

\epsfig{file=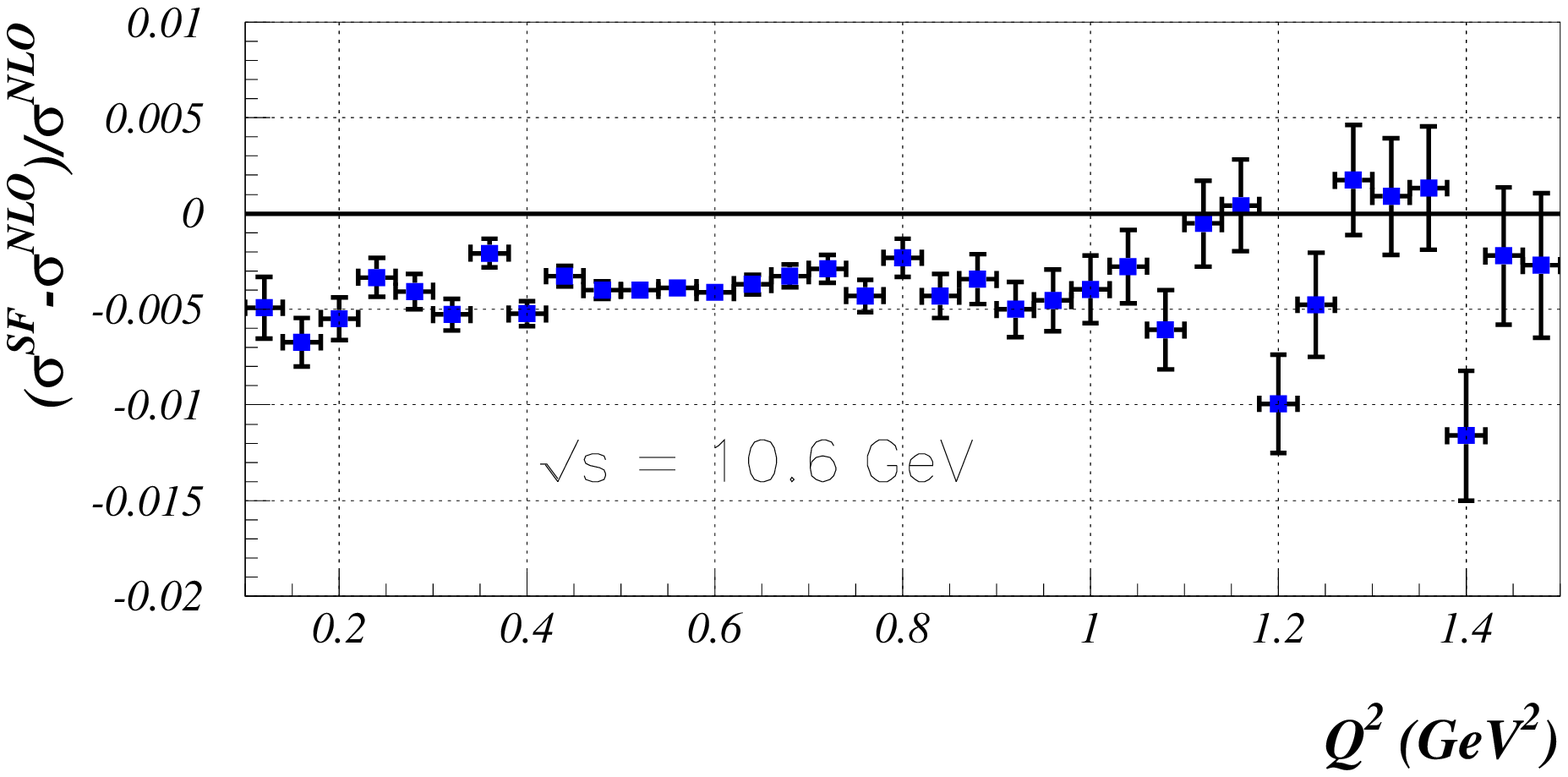,width=7.5cm}
\end{center}
\caption{Comparison between the collinear approximation
by structure functions and the fixed-order NLO result
for different cms energies. Cuts from Table~\ref{tab:cuts}.}
\label{fig:collnloc}
\end{figure}

The original and default version of EVA~\cite{Binner:1999bt}, simulating 
the process $e^+ e^- \rightarrow \pi^+ \pi^- \gamma$ at LO, allowed for 
additional initial-state radiation of soft and collinear photons by the 
structure-function method~\cite{Caffo:1997yy}. By convoluting 
the Born cross-section with a given SF distribution,
soft photons are resummed to all orders in perturbation theory 
and large logarithms of collinear origin, $L=\log(s/m_e^2)$, are taken 
into account up to two-loop approximation. 
The NLO result, being a complete one-loop result, contains these logarithms 
in order $\alpha$ and additional subleading terms, which of course
are not taken into account within the SF method. Although small, 
these subleading contributions become important 
for measurements aiming at an accuracy better than 1$\%$. 
Furthermore, a NLO generator is more suitable for comparison with an
experimental setup. While in the SF approach the extra collinear photon 
emission is integrated out and some information is lost,
in a fixed-order calculation the full angular dependence is kept
and energy-momentum is conserved, i.e. the sum of the momenta 
of the generated outgoing particles agrees with the incoming 
$e^+ e^-$ momentum. 

Table~\ref{tab:xsec} presents the total cross-section 
for $e^+ e^- \to \pi^+ \pi^- \gamma (\gamma)$ calculated at LO 
and NLO for three different cms energies with the kinematical cuts 
listed in Table~\ref{tab:cuts}.
Two NLO predictions are shown. The first one, NLO(1), which can be compared 
with the SF result derived from EVA, includes a cut on the invariant 
mass of the hadrons plus the tagged photon. The last was introduced
in~\cite{Binner:1999bt} in order to reduce the kinematic distortion of 
the events due to the extra collinear emission.
The second prediction, NLO(2), is obtained 
without this cut. The $Q^2$ dependence of the difference 
between the NLO(1) prediction and the SF result is showed 
in Fig.~\ref{fig:collnloc}.
The size and sign of the NLO corrections do depend on the particular choice 
of the experimental cuts. Hence only using a Monte Carlo event generator can 
one realistically compare theoretical predictions with experiment and 
extract \(R(s)\) from the data. 

The results of EVA and those denoted NLO(1) for the total 
cross-section are in reasonably good agreement. Both of them are 
clearly sensitive to the cut on $M^2_{\pi^+ \pi^- \gamma}$.
This cut dependence is displayed in Table~\ref{tab:invariantcut}.
Remarkably, the typical difference between the results of the 
two programs is clearly less than $0.5\%$ for most of the entries.

The systematic uncertainty of the program, due to inadequate 
treatment of ISR, namely missing
leading logarithmic higher order corrections and lepton pair production, 
is conservatively estimated to be of around $0.5\%$ in the total 
cross-section~\cite{Rodrigo:2001kf}.

\section{MUON PAIR PRODUCTION}

\begin{figure*}
\vspace{-.8cm}
\begin{center}
\epsfig{file=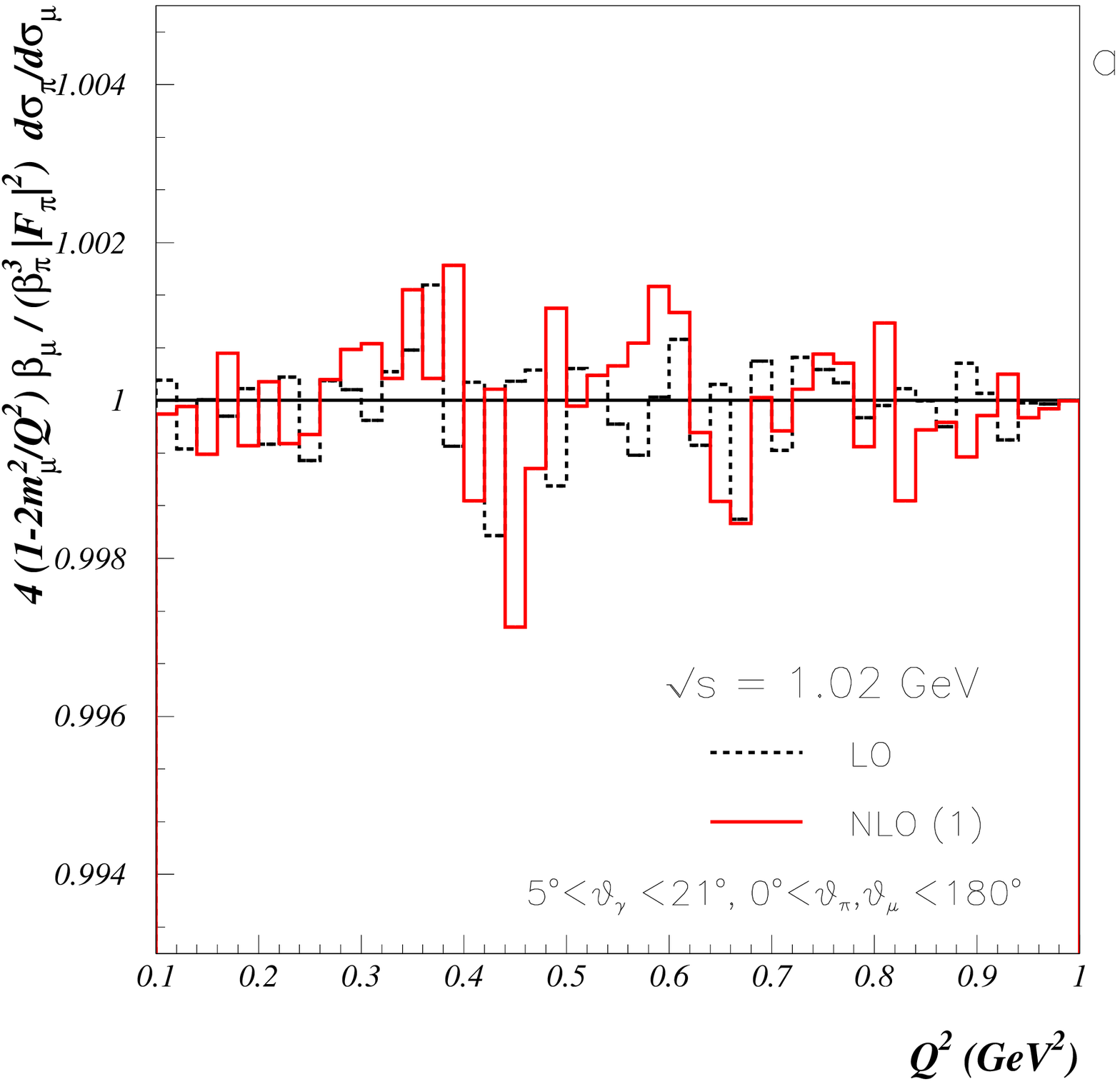,width=7cm}
\epsfig{file=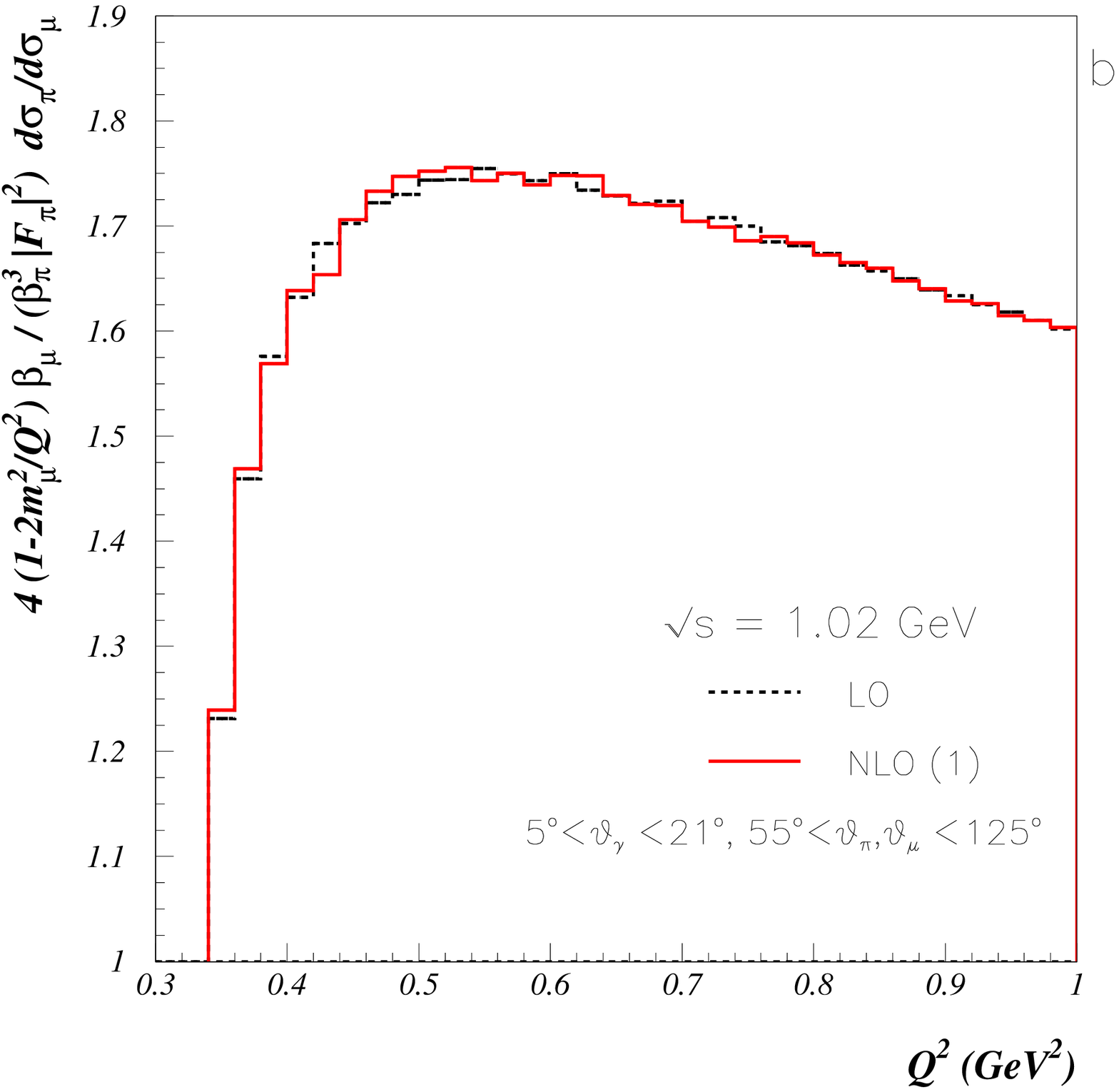,width=7cm}  \vspace{-1cm}
\end{center}
\caption{Ratio between pion and muon yields, after dividing through 
their respective R-ratio. a: no cuts on pion and muon angles;
b: with angular cuts on pion and muon angles.}
\label{fig:pipimumu}
\end{figure*}

Results for the total cross-section of muon pair production 
are listed in Table~\ref{tab:mxsec}. The radiative muon 
cross-section can be used for a calibration of the pion yield.
A number of radiative corrections are expected to cancel in the 
ratio. For this reason we consider the ratio between 
the pion and the muon yields, after dividing the 
former by $|F_{\pi} (Q^2)|^2 (1-4m_\pi^2/Q^2)^{3/2}$, 
the latter by  $4 (1+2m_{\mu}^2/Q^2) \sqrt{1-4m_{\mu}^2/Q^2}$. 
In Fig.~\ref{fig:pipimumu}a we consider the full angular range for pions 
and muons, with $\theta_\gamma$ between $5^\circ$ and $21^\circ$.
Clearly all radiative corrections and kinematic effects disappear,
up to statistical fluctuations, in the leading order program as well 
as after inclusion of the NLO corrections.

In Fig.~\ref{fig:pipimumu}b an additional cut on pion and muon angles 
has been imposed. As demonstrated in Fig.~\ref{fig:pipimumu}b, the ratio 
differs from unity once (identical) angular cuts are imposed on pions 
and muons, a consequence of their different angular distribution.
To derive the pion form factor from the ratio between pion and muon 
yields, this effect has to be incorporated. However, the correction 
function shown in Fig.~\ref{fig:pipimumu}b is independent from the 
form factor, and hence universal and model-independent (ignoring FSR 
for the moment).

\section{UNTAGGED PHOTONS}

\begin{figure}
\begin{center}
\epsfig{file=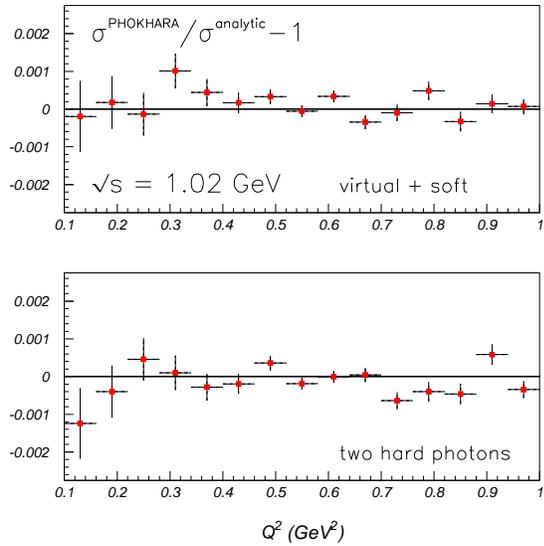,width=8cm} \vspace{-1.5cm}
\end{center}
\caption{Comparison of the virtual+soft and hard contributions to the 
$\pi^+ \pi^-$ differential cross-section with inclusive analytical results.
Soft photon cutoff: $w=10^{-4}$.}
\label{fig:realvirt}
\end{figure}

Both EVA and PHOKHARA were initially designed to simulate reactions with 
tagged photons, i.e. at least one photon was required to be emitted 
under large angles. The extension of these results to untagged 
photon events, i.e. for photons emitted at arbitrary small angles,
was recently investigated~\cite{Kuhn:2002xg,Rodrigo:2001cc}.

An important ingredient in the extension of the NLO Monte Carlo
program PHOKHARA to small photon angles is the evaluation of the virtual
corrections to reaction (\ref{eq:reaction}) in the limit $m_e^2/s \ll 1$,
which are equally valid for large and small angles.  
Compact results for the one-loop two-, three- and four-point functions 
that enter this calculation can be found in the 
literature~\cite{'tHooft:1978xw,Beenakker:1988jr} for arbitrary
values of $m_e^2/s$. However, the combination of these analytical
expressions with the relevant coefficients is numerically unstable in
the limit of small mass and angles. A compact, numerically stable result, 
valid for an arbitrarily small photon angle, is therefore required.
As a consequence of the highly singular kinematic coefficients, terms
proportional to $m_e^2$ and even $m_e^4$  must be kept in the expansion, 
which will contribute, after angular integration, to the total 
cross-section even in the limit $m_e^2/s\to0$.  

Inclusive NLO calculations can be used to test the performance of 
the simulation at small angles. Figure~\ref{fig:realvirt} shows the 
comparison of the differential cross-section 
for the $\pi^+ \pi^-$ mode with the analytical results
from~\cite{Berends:1986yy,Berends:1988ab}. The agreement 
is excellent.

\section{ISR DOMINANCE AND FSR}

\begin{figure}
\begin{center}
\vspace{-.8cm}
\epsfig{file=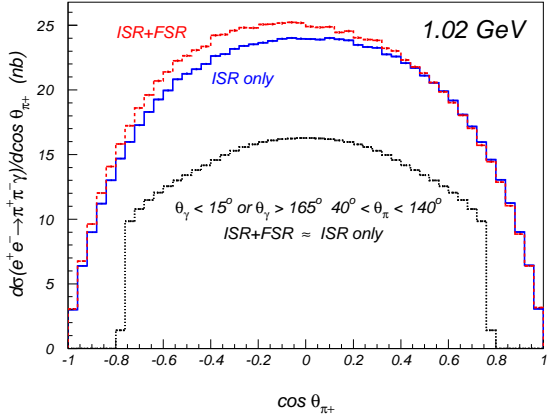,width=8cm} \vspace{-1.5cm}
\end{center}
\caption{Angular distribution of $\pi^+$ with and without FSR 
for different angular cuts.}
\label{fig:angular}
\end{figure}

\begin{figure}
\begin{center}
\vspace{-.8cm}
\epsfig{file=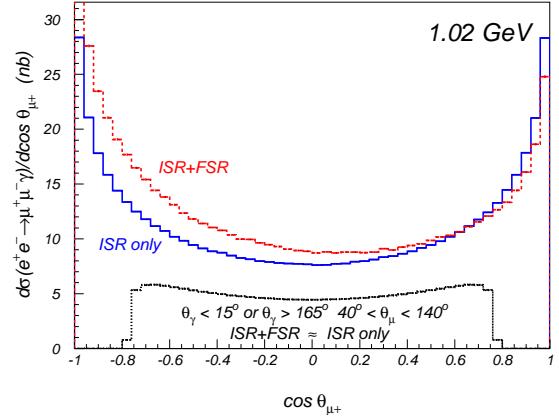,width=8cm} \vspace{-1.5cm}
\end{center}
\caption{Angular distribution of $\mu^+$ with and without FSR 
for different angular cuts.}
\label{fig:mangular}
\end{figure}

Final-state radiation should be regarded as the background of the 
measurement and would be required for the complete
simulation of the reaction. The complete leading order matrix element 
squared is given by 
\begin{equation}
|{\cal M}|^2 = |{\cal M}_{\mathrm{ISR}}|^2  + |{\cal M}_{\mathrm{FSR}}|^2
+2 \mathrm{Re}[{\cal M}_{\mathrm{ISR}} {\cal M}_{\mathrm{FSR}}^\dagger]~.
\label{eq:interference}
\end{equation}
FSR and its interference with ISR were already included in 
EVA~\cite{Binner:1999bt} for the two-pion case. 
The pions were assumed to be point-like, and scalar QED was 
applied to simulate photon emission off the charged pions. 
There it was demonstrated that suitably chosen configurations, namely 
those with hard photons at small angles relative to the beam and well 
separated from the pions, are dominated by ISR. FSR can therefore 
be reduced to a reasonable limit or be controlled by the simulation.

At B-factories, where one has to deal with very hard tagged photons,
the situation is even better because the kinematic separation between the 
photon and the hadrons becomes very clear.
For events where hadrons and photon are produced mainly back to back 
the suppression of FSR is naturally accomplished.

\begin{figure*}
\begin{center}
\epsfig{file=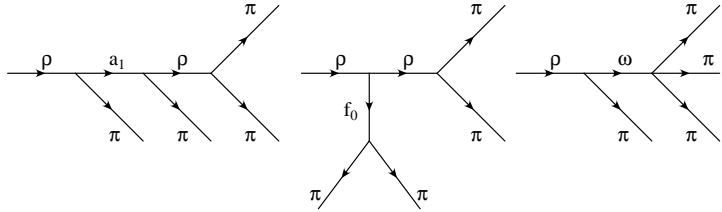,width=9.5cm} \vspace{-.5cm}
\end{center}
\caption{Diagrams contributing to the 4$\pi$ hadronic current.}
\label{fig:4picurrent}
\end{figure*}

The third term in the right-hand side of Eq.~(\ref{eq:interference}),
the ISR--FSR interference, is odd under charge conjugation and its
contribution vanishes under angular integration.
It gives rise, however, to a relatively large charge asymmetry and, 
correspondingly, to a forward--backward asymmetry
\begin{equation}
A(\theta) = \frac{N^{\pi^+}(\theta)-N^{\pi^+}(\pi-\theta)}
{N^{\pi^+}(\theta)+N^{\pi^+}(\pi-\theta)}~.
\end{equation}
The asymmetry can be used for calibration of the FSR amplitude,
and fits to the angular distribution \(A(\theta)\) can test
details of its model dependence.

This is illustrated in Figs.~\ref{fig:angular} and~\ref{fig:mangular},
where the angular distributions of $\pi^+$ and $\mu^+$ respectively are 
shown for different kinematical cuts. The angles are defined with respect 
to the incoming positron. If no angular cut is applied 
the angular distribution in both cases is highly asymmetric as 
a consequence of the ISR--FSR interference contribution. If cuts 
suitable to suppress FSR, and therefore the ISR--FSR interference,
are applied the distributions become symmetric. These plots have
been obtained with the new version of PHOKHARA, which now incorporates at LO
FSR, and its interference with ISR, for two pions (point-like) and muons.

\section{NEW HADRONIC CHANNELS}

Because of the modular structure of PHOKHARA additional hadronic 
modes can be easily implemented. The four-pion channels
($2\pi^+ 2\pi^-$ and $2\pi^0\pi^+\pi^-$), which give the dominant 
contribution to the hadronic cross-section in the region from 1 to 2~GeV,
are a new feature of our event generator. 

Isospin invariance relates the amplitudes of the 
$e^+e^- \to 2\pi^+ 2\pi^-$ and $e^+e^- \to 2\pi^0\pi^+\pi^-$ processes 
and those for $\tau$ decays into $\pi^-3\pi^0$ and $\pi^+2\pi^-\pi^0$.
The description of the four-pion hadronic current 
follows~\cite{Czyz:2000wh,Decker:1994af}. The basic building blocks
of this current are schematically depicted in Fig.~\ref{fig:4picurrent}
and described in detail in~\cite{Czyz:2000wh}.

\begin{figure}
\begin{center}
\epsfig{file=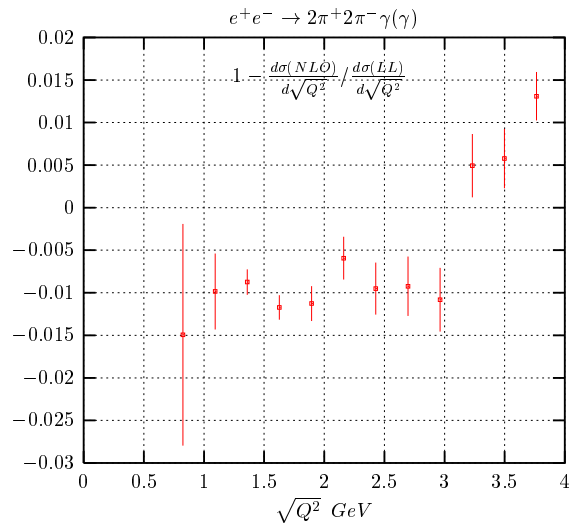,width=7.5cm} \vspace{-1.5cm}
\end{center}
\caption{Relative non-leading contribution to the four pion
differential cross-section at $\sqrt{s}$=4~GeV.}
\label{fig:4pillnlo}
\end{figure}

Results obtained with PHOKHARA for this channel have been 
compared with the Monte Carlo, which simulates the same 
process at LO~\cite{Czyz:2000wh} and includes additional 
collinear radiation through SF techniques. Typically, differences 
of order 1$\%$ are found (see Fig.~\ref{fig:4pillnlo}), which are of the 
expected size and of the same order as for the two-pion final 
state~\cite{Rodrigo:2001kf}. Details of the comparison and of the 
performed tests will be presented in a forthcoming 
publication~\cite{inpreparation}.

The implementation of the three-pion mode is under way. Other hadronic 
channels will be likewise considered in the future. 

\section{CONCLUSIONS}

The PHOKHARA Monte Carlo event generator has been upgraded.
It now includes, besides the $\pi^+ \pi^-$ and $\mu^+ \mu^-$ 
channels, also $2\pi^+ 2\pi^-$ and $2\pi^0\pi^+\pi^-$ as final states.
FSR, and its interference with ISR, have also been included 
for the first two channels. Furthermore, the simulation of 
events where the photon(s) is emitted under small angles 
with respect to the beam axis has been tested numerically. 
Further upgrades are in progress.

\section*{ACKNOWLEDGEMENTS}

It is a pleasure to thank the organizers of this meeting for the 
stimulating atmosphere created during the workshop. 
Special thanks to S.~Vascotto for carefully reading the manuscript. 
Work supported in part by BMBF under grant number 05HT9VKB0, 
E.U. EURIDICE network RTN project HPRN-CT-2002-00311,  
TARI under contract HPRI-CT-1999-00088, and  
MCyT, Plan Nacional I+D+I (Spain) under grant BFM2002-00568.


\end{document}